\documentclass[]{spie}  
\usepackage{booktabs}
 
\usepackage{amsmath,amsfonts,amssymb}
\usepackage{graphicx}
\usepackage[colorlinks=true, allcolors=blue]{hyperref}
\usepackage{cite}
\title{A deep learning approach for patchless estimation of ultrasound quantitative parametric image\\ with uncertainty measurement}

\author[a]{Ali K. Z. Tehrani}
\author[b]{Ivan M. Rosado-Mendez}
\author[b]{Hayley Whitson}
\author[a]{Hassan Rivaz}
\affil[a]{Department of Electrical and Computer Engineering, Concordia University, Canada.}
\affil[b]{Department of Medical Physics and Radiology, University of Wisconsin, United States.}

\authorinfo{Further author information: (Send correspondence to H. Rivaz)\\H. Rivaz: E-mail: hrivaz@ece.concordia.ca}

\begin{document} 
\maketitle

\begin{abstract}
Quantitative ultrasound (QUS) aims to find properties of scatterers which are related to the tissue microstructure. Among different QUS parameters, scatterer number density has been found to be a reliable biomarker to detect different abnormalities. The homodyned K-distribution (HK-distribution) is a model for the probability density function of the ultrasound echo amplitude that can model different scattering scenarios but requires a large number of samples to be estimated reliably. Parametric images of HK-distribution parameters can be formed by dividing the envelope data into small overlapping patches and estimating parameters within the patches independently. This approach imposes two limiting constraints: the HK-distribution parameters are assumed to be constant within each patch, and each patch requires enough independent samples. In order to mitigate those problems, we employ a deep learning approach to estimate parametric images of scatterer number density (related to HK-distribution shape parameter) without patching. Furthermore, an uncertainty map of the network’s prediction is quantified to provide insight about the confidence of the network about the estimated HK parameter values. 
\end{abstract}


\section{INTRODUCTION}
\label{sec:intro}  
Quantitative ultrasound (QUS) finds scatterers' properties which are highly related to the tissue microstructure \cite{oelze2016review}. Different QUS parameters such as speed of sound, scatterer number density, and backscattering coefficient have been employed for characterization of tissues. Among different QUS parameters, scatterer number density has been found to be useful for different clinical applications such as liver fibrosis detection \cite{zhou2020value} and Hepatic steatosis assessment \cite{fang2020ultrasound}. Scatterer number density can be quantified by modeling the ultrasound echo amplitude using different probability density functions. The homodyned K-distribution (HK-distribution) is a well-known distribution that has been widely used to quantify the scatterer number density and characterizing the tissues. This distribution can model diverse scattering scenarios but requires a large number of independent samples \cite{destrempes2013estimation}.

The HK-distribution parametric images are formed by dividing the envelope data into small overlapping patches and estimating parameters within the patches independently. Estimating the parametric images is challenging  and the parametric images are often noisy but still have been found useful in clinical applications to detect different abnormalities \cite{tsui2008performance}. The main challenge in estimating the HK-distribution parametric images is the small size of the patches. Increasing the size of the patches might not be helpful since the spatial resolution will be lost and the heterogeneity inside the patch is increased (due to heterogeneity of tissue types or spatially variant nature of point spread function). Correlated samples also should be skipped since they introduce bias to the estimation of HK-distribution parameters. Therefore, a large number of samples inside each patch should be ignored to reduce the correlation between the samples which further reduces the number of samples.   
          

In this paper, an ultrasound simulation method is employed to generate the training data. Unlike sampling from HK-distribution, this method of simulation data generation contains correlated samples which is a more realistic method and closer to the real ultrasound data \cite{tehrani2022robust}. In addition to this, multiple frames from experimental phantom data are employed to obtain a more accurate parametric images and quantify the uncertainty map of the network. The uncertainty map would enable the clinicians to find out how reliable the estimations are in each region.     

\section{Materials and Methods}

\subsection{HK-distribution}
The HK-distribution can be formulated as \cite{destrempes2013estimation}:
\begin{equation}
	P_{HK}(A|\varepsilon ,\sigma^2,\alpha) = A\int_{0}^{\infty}uJ_{0}(u\epsilon)J_{0}(uA)(1+\frac{u^2\sigma^2}{2})^{-\alpha}du
\end{equation}
where $A$ denotes the envelope of the backscattered echo ultrasound data, $u$ is the variable of integral that needs to be integrated from zero to infinity, $J_{0}(.)$ is the zero-order Bessel function, and $\alpha$ is the scatterer clustering parameter which reflects to the scatterer number density. The parameter $\alpha$ and the ratio of coherent signal power ($\epsilon^2$) to the diffuse one ($2\sigma^2\alpha$) denoted as $k$ have been employed as the parameters of HK-distribution to characterize the tissues.  
\subsection{Data Generation}
\begin{figure}[!t]
	\centering
	\includegraphics[width=0.97\textwidth]{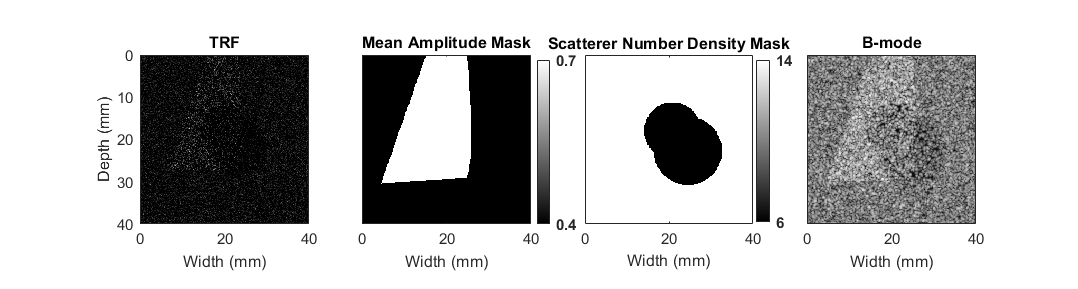}
	\centering
	\caption{One example of generated simulation data.} 
	\label{fig:sample}
\end{figure}

A diverse dataset with known scatterer number density is required to train the network. We followed the fast grid-based method \cite{tehrani2022robust} with the difference that here the scatterer number density can be any value in the range of 1-20. Assuming   weak   scattering, the ultrasound RF data can be obtained by the 2D convolution of Point Spread Function (PSF) and the Tissue Reflectivity Function (TRF) \cite{zhang2020deep,tehrani2022robust}.
 \begin{equation}    
   s_{(a,l)} = TRF_{(a,l)} \ast h_{(a,l)}
\end{equation}
Where $h$ denotes the PSF. In order to have a known scatterer number density, the PSF is assumed to be spatially invariant which can be described as:
\begin{equation}
	h[a,l]=e^{-\frac{1}{2}(\frac{a^2}{{\sigma_{a}}^{2}}+ \frac{l^2}{{\sigma_{l}}^{2}})} \times cos(2\pi f_{c}a)  
\end{equation}
where the PSF is modeled as a 2D Gaussian modulated with a cosine with the center frequency of $f_c$ in the axial direction. The parameters ${\sigma_{a}}^{2}$, and ${\sigma_{l}}^{2}$ denote the axial and lateral width of the Gaussian function and they are related to the resolution cell.  
The TRF is the 2D map that contains all point scatterers. In each grid point, only one scatterer is allowed to be present and its amplitude is sampled from normal distribution. The mean of the normal distribution is randomly selected from values 1 to 5, the variance is fixed to 0.02 of the mean value. The TRF is constructed using shapes having different values of scatterer number density and mean scattering amplitudes. One sample generated by this method is shown in Fig. \ref{fig:sample}. We generated 11000 samples to train the network. It should be noted that the ground truth value of scatterer number density is known and depends on how the resolution cell size is defined. Here, we define the resolution cell in each direction as $3\times \sigma_x$, where x is the axial or lateral directions. Different definition of the resolution cell results in different value of the scatterer number density. Please refer to \cite{tehrani2022robust} for more information about the grid-based simulation method. In order to reduce the correlation, samples are skipped from both axial and lateral directions to have an image of size $256 \times 128$, and an average correlation between samples of 0.28 is obtained.

\subsection{Conventional Patch-based Methods}
 Hruska \textit{et al.} employed SNR, skewness, and kurtosis to estimate HK-distribution parameters \cite{Hruska2009}. Destrempes \textit{et al.} proposed XU estimator in which two log moment statistics (named as X and U) were employed, and reported improved parameter estimation \cite{destrempes2013estimation}. The X, and U can be defined as:
 \begin{equation}
 	\begin{gathered}
 		X = <Ilog(I)>/<I> - <log(I)>,\\
 		U = <log(I)>-log(<I>),
 	\end{gathered}
 \end{equation} 
where $<.>$ denotes sample mean, and $I=A^2$ is the intensity of backscattered signal. The XU estimator iteratively solves a constraint optimization using bisection method to find the parameters. We employed this method for comparison.  
\subsection{Experimental Phantom Data}
Several experimental phantoms have been employed to evaluate the method.

1) Layered Phantom: A phantom with a middle layer having different properties than the top and bottom layers reported in previous publication \cite{nam2012comparison} has been employed. The phantom was constructed by an emulsion of ultrafiltered milk and water-based gelatin, and 5–43 $\mu m$ diameter glass beads were utilized as the source of scattering. A Siemens Acuson S2000 scanner (Siemens Medical Solutions USA, Inc.) with 18L6 probe linear transducer having the center frequency of 8.9 MHz was used for data collection. The layer with higher intensity has a higher backscattering coefficient and scatterer concentration. The backscattering coefficient of this layer is $6.37 \times 10^{-3}$ $cm^{-1} sr^{-1}$ and it is $3.52 \times 10^{-3}$  $cm^{-1}$ $sr^{-1}$ for other parts at the center frequency.

2) CIRS Phantom: Data from a multipurpose CIRS phantom (model 040GSE, Norfolk, VA, USA) was collected at Concordia university using E-CUBE 12 Alpinion machine by an L3-12H transducer with a sampling rate of 40 MHz and center frequency of 8.5 MHz.

3) Gammex Phantom: Data was collected at university of Wisconsin using a Verasonics Vantage 128 System (Verasonics,Kirkland,  WA,  USA) machine with an L11-5v transducer operating at 8 MHz. 

\subsection{Pre-Processing of experimental phantom data} 
\begin{figure}[!t]
	\centering
	\includegraphics[width=0.97\textwidth]{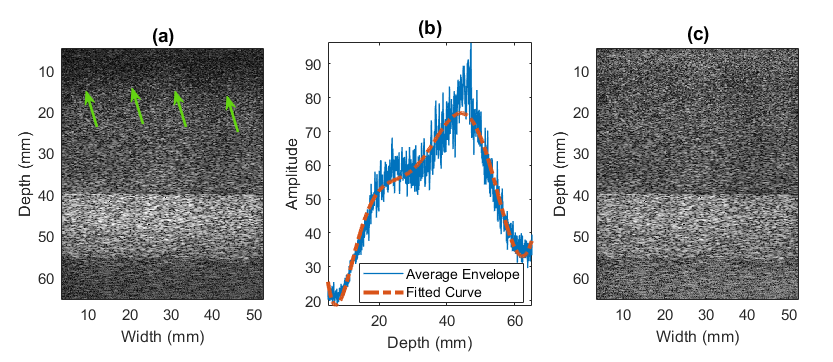}
	\centering
	\caption{Layered phantom before the system effect compensation (a). The envelope amplitude average of reference phantoms and the fitted curve (b). The compensated layered phantom (c). The intensity variation of the phantom before compensation is clear and the area with lower intensity is marked. The phantom data has been reported in previous publication \cite{nam2012comparison}.} 
	\label{fig:comp}
\end{figure}

Experimental phantom data may contain variations of intensity due to beamforming, focusing and time gain control (TGC) which are not present in the simulation data and leads to performance decay of parameter estimation. Inspired by the reference phantom method which cancels out the system effect \cite{Rosado2016}, we employ an intensity normalization technique to reduce the system effect on the envelope data intensity. First, the average of envelope amplitudes in each depth is computed across several frames of uniform reference phantoms imaged by the same imaging settings. In the next step, a curve is fitted to the obtained average envelope amplitude. The inverse of the fitted curve can be utilized to normalize the experimental phantom data imaged by the same settings. The procedure is depicted in the Fig. \ref{fig:comp}. It can be observed that the normalized data has a more uniform intensity across different depth compared to the pre-normalized one. 
\subsection{Multi-frame acquisition and data processing for experimental phantoms}  
For the layered phantom, the probe is moved in the out-of-plane direction to collect 12 frames from this phantom. We skipped 7 samples in the axial and 1 sample in the lateral direction from the acquired envelope data to reduce the correlation from 0.75 (no skipping) to 0.141. The patch size for XU algorithm is $4.5\times 4.5$ $mm$ with $75 \%$ overlap.    

For the CIRS phantom, the probe is moved in the out-of-plane direction to collect 10 frames from this phantom. The patch size for XU algorithm is $4.5\times 4.5$ $mm$ with $75 \%$ overlap. We skipped 4 samples in the axial and 0 sample in the lateral direction from the acquired envelope data to reduce the correlation from 0.95 (no skipping) to 0.49. In another experiment, we skipped 5 samples in the axial and 1 sample in the lateral direction to have the correlation of 0.36 to see the effect of different correlations. It should be noted that skipping more samples results in reduction of the bias but lower number of samples would be available to estimate the scatterer number density which makes estimation of small parts challenging.  

For the Gammex phantom, the probe is moved in the out-of-plane direction to collect 12 frames from this phantom. The patch size for XU algorithm is $4.0\times 4.0$ $mm$ with $75 \%$ overlap. We skipped 6 samples in the axial and 1 sample in the lateral direction from the acquired envelope data to reduce the correlation from 0.95 (no skipping) to 0.55. In another experiment, we skipped 13 samples in the axial and 1 sample in the lateral direction to have the correlation of 0.19 to see the effect of different correlations.    
\subsection{Prediction and uncertainty quantification }        
For experimental phantom data, several frames are collected by sweeping the probe in the out-of-plane direction. The final estimate and its corresponding uncertainty can be quantified as: 
 	\begin{equation}
 	\begin{gathered}
 \widetilde{S}= \frac{1}{N_f}\sum_{i=1}^{N_f}\widetilde{S_i}, \\
 Uncertainty=\frac{\sqrt{ \frac{1}{N_f}\sum_{i=1}^{N_f}(\widetilde{S_i}- \widetilde{S})^2}}{\widetilde{S}},
 	\end{gathered}
    \end{equation}  
where the final estimate ($\widetilde{S}$) is calculated by averaging the estimates ($\widetilde{S_i}$) obtained from $N_f$ frames. The proposed uncertainty measures how different the estimates are across different frames. If the estimates from all frames are similar, the uncertainty will be low. On the contrary, high uncertainty is expected when the estimates are different. It should be noted that this is a frame-wise uncertainty measurement meaning that if there is no difference in the estimated value across different frames, the uncertainty is low even if the estimated values are incorrect.   
\subsection{Network architecture and training}
We employed Deeplab V3 which has shown excellent performance in semantic segmentation \cite{yurtkulu2019semantic}. The last layer of the model is altered to have one output as the scatterer number density. The following loss function is employed:

\begin{equation}
loss = \frac{1}{N} \sum_{}^{} (log_{10}(\widetilde{S}) - log_{10}(S))^2
\end{equation}
where $\widetilde{S}$ and $S$ are the predicted and ground truth scatterer number density, respectively. The log compression is employed to avoid the bias toward high scatterer number density values. Rectified Adam (RAdam) is employed as the optimizer which has shown outperforming Adam and be robust to the selection of learning rate \cite{liu2019variance}. The network is trained on single NVIDIA RTX 3090 with 24 GB of memory.

\section{Results}
\subsection{Simulation Results}
\begin{figure}[!t]
	\centering
	\includegraphics[width=0.98\textwidth]{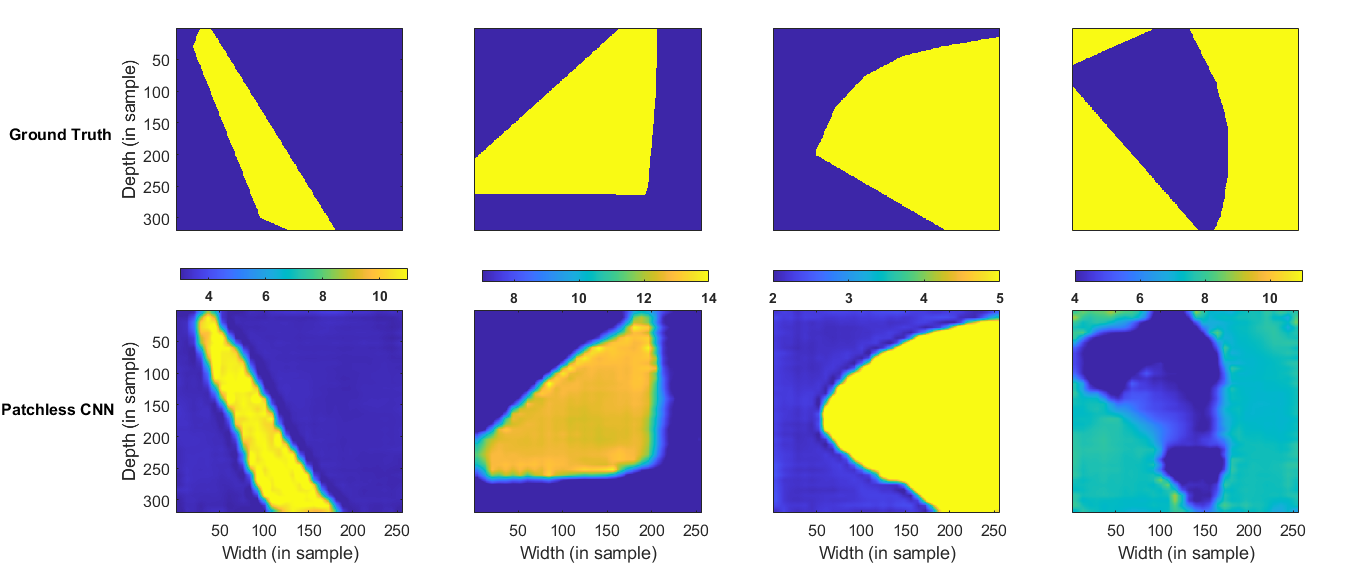}
	\centering
	\caption{Four examples of the patchless CNN estimation of scatterer number density. The parametric image values are underestimated in the forth sample.} 
	\label{fig:sim_res}
\end{figure}

The method is evaluated on 1000 test data generated in a similar way of training data. Due to the fact that the value of the parameter $\alpha$ is not known for the simulation data, we cannot compare it with the XU estimator. The results are given in Table \ref{tab:sim}. The metrics are evaluated on each sample of test dataset and, the average and standard deviation are reported. Four samples of the simulation results are depicted in Fig. \ref{fig:sim_res}.
\begin{table}[ht]
	\centering
	\caption{The simulation results of the proposed method.}
	\label{tab:sim}
	\begin{tabular}{@{}cccc@{}}
		\toprule
		& RMSE          & RRMSE         & MAE           \\ \midrule
		Patchless CNN & 1.12$\pm$0.59 & 0.42$\pm$0.19 & 0.85$\pm$0.55 \\ \bottomrule
	\end{tabular}
\end{table}
\subsection{Experimental phantom results}
\subsubsection{Layered phantom results}  
\begin{figure}[!t]
  	\centering
  	\includegraphics[width=0.99\textwidth]{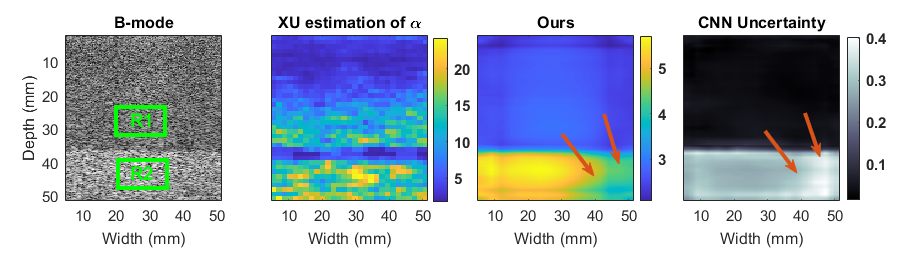}
  	\centering
  	\caption{From left to right: B-mode image, estimated $\alpha$ by XU algorithm, patchless CNN, and uncertainty. The area that patchless CNN incorrectly underestimated the scatterer number density is marked and uncertainty is higher in that area.  } 
  	\label{fig:layered_res}
  \end{figure}

The predicted results of the CNN and XU are the average across the collected frames. The uncertainty is measured by dividing the standard deviation of the results by their mean value. The results of the patchless CNN and XU algorithm are given in Fig. \ref{fig:layered_res}. The uncertainty map has a higher value in the region where the CNN incorrectly estimates a lower value in the phantom (the region is marked with red arrows). 

In order to quantitatively investigate the presented results, the ratio of the mean of obtained values in regions R2 and R1 (highlighted in the B-mode image) are computed and compared with the ratio of their known backscattering coefficient. The ratio is $\frac{6.37}{3.52} = 1.81$ for the backscattering coefficient, it is $\frac{16.34\pm3.72}{10.99\pm 2.73}=1.61\pm 0.63$ for XU, and $\frac{5.19\pm0.31}{1.81\pm 0.03}=1.81 \pm 0.11$ for the patchless CNN. It can be seen that the ratio for the patchless CNN is closer to the true backscattering coefficient ratio compared to the XU method which demonstrates that this method has a lower bias compared to XU.      
\subsubsection{CIRS phantom results}

\begin{figure}[!t]
 	\centering
 	\includegraphics[width=0.97\textwidth]{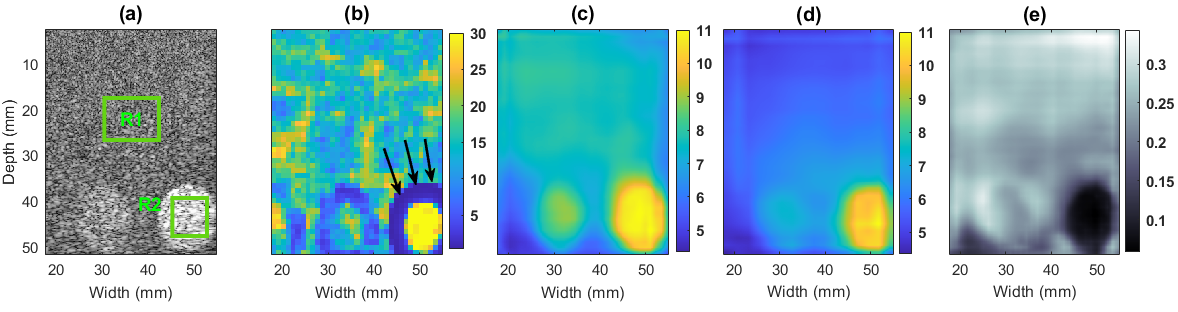}
 	\centering
 	\caption{Results of CIRS phantom. B-mode image (1), estimated $\alpha$ by XU algorithm (b), patchless CNN using samples with correlation 0.49 (skip 4 samples in axial direction) (c), patchless CNN using samples with correlation 0.36 (skip 5 samples in axial and 1 in the lateral direction) (d), and uncertainty obtained from c (e).} 
 	\label{fig:cirs_res}
 \end{figure}

The results are shown in Fig. \ref{fig:cirs_res}. The XU algorithm fails to obtain reliable values on the borders (marked in b) of the inclusions due to presence of two different distributions inside the patches of those regions, while the CNN results do not suffer from this issue. The CNN provides higher contrast and less bias with the samples with lower correlation, but due to skipping more samples, the inclusion borders are not as clear as with the higher correlation one. This demonstrates a trade off between skipping the samples to reduce the correlation (lower bias) and spatial resolution. This issue was less identifiable in the layered phantom since there were adequate samples in each layer; therefore, we were able to skip enough samples to reduce the bias.     
\subsubsection{Gammex phantom results}

\begin{figure}[!t]
	\centering
	\includegraphics[width=0.97\textwidth]{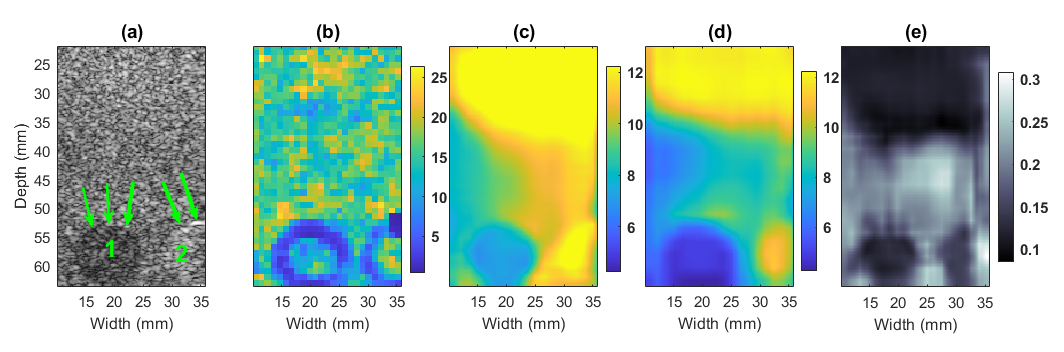}
	\centering
	\caption{Results of Gammex phantom. B-mode image (1), estimated $\alpha$ by XU algorithm (b), patchless CNN using samples with correlation 0.55 (skip 6 samples in axial and 1 in lateral direction) (c), patchless CNN using samples with correlation 0.19 (skip 13 samples in axial and 1 in the lateral direction) (d), and uncertainty obtained from c (e).} 
	\label{fig:gammex_res}
\end{figure}

The results are shown in Fig. \ref{fig:gammex_res}. The parametric image obtained by XU algorithm has artifacts around the inclusions similar to Fig. \ref{fig:cirs_res}. The parametric image with higher correlation (c) has higher spatial resolution but it has more biased values (compare the inclusion 1 in (c) and (d)). The inclusion 1 is clearly detected by the CNN for both inputs (c and d), but in (c) the inclusion has a more biased values than (d) due to presence of higher correlation. The inclusion 2 is also hardly detected by the CNN since only part of it is available.

\section{Discussion}
In this paper, a deep learning frame work is proposed to quantify the scatterer number density. One important thing to note is that the method of calculating the scatterer number density in simulation data generation dictates the output of the network. In other word, different definition of resolution cell in simulation data generation can change the ground truth value which also alters the output of the network.

Another crucial point is that the correlation between samples can affect the results. Most of recent work focus on estimating the scatterer number density using training data obtained from i.i.d samples. Although these samples contain more information than the correlated ones, achieving this condition in real data is difficult and requires skipping many samples which is not efficient in parametric image reconstruction. In this work, the approach is changed by generating simulation data using the grid-based method to have correlation between the samples. The disadvantage of this approach is that the exact value of $\alpha$ is not known, and generating coherent component is complex. 

The impact of different correlations for the experimental phantoms is also investigated. Generally, lower correlation can help having less bias in estimation but achieving this might cause losing spatial resolution due to requiring skipping many samples.

\subsection{Conclusion}
In this paper, a patchless deep learning solution was proposed to obtain scatterer number density parametric images. The method was validated using simulation data and three experimental phantoms. Frame-wise uncertainty map was also obtained from the parametric images of the multiple frames. In order to enable the network to predict the experimental phantom data having correlated samples, a grid-based simulation in which correlated samples were present was employed for training data generation. The proposed method was trained on simulation data and was able to reconstruct scatterer number density parametric image of experimental phantoms imaged by different scanners.    

\subsection{Acknowledgments}
We acknowledge the support of the Natural Sciences and Engineering
Research Council of Canada (NSERC), and the report of the layered phantom data in the previous publication \cite{nam2012comparison}.

\bibliography{refs3} 

\begin{thebibliography}{10}

\bibitem{oelze2016review}
Oelze, M.~L. and Mamou, J., ``Review of quantitative ultrasound: Envelope
  statistics and backscatter coefficient imaging and contributions to
  diagnostic ultrasound,'' {\em IEEE transactions on ultrasonics,
  ferroelectrics, and frequency control}~{\bf 63}(2),  336--351 (2016).

\bibitem{zhou2020value}
Zhou, Z., Fang, J., Cristea, A., Lin, Y.-H., Tsai, Y.-W., Wan, Y.-L., Yeow,
  K.-M., Ho, M.-C., and Tsui, P.-H., ``Value of homodyned k distribution in
  ultrasound parametric imaging of hepatic steatosis: An animal study,'' {\em
  Ultrasonics}~{\bf 101},  106001 (2020).

\bibitem{fang2020ultrasound}
Fang, F., Fang, J., Li, Q., Tai, D.-I., Wan, Y.-L., Tamura, K., Yamaguchi, T.,
  and Tsui, P.-H., ``Ultrasound assessment of hepatic steatosis by using the
  double nakagami distribution: a feasibility study,'' {\em Diagnostics}~{\bf
  10}(8),  557 (2020).

\bibitem{destrempes2013estimation}
Destrempes, F., Por{\'e}e, J., and Cloutier, G., ``Estimation method of the
  homodyned k-distribution based on the mean intensity and two log-moments,''
  {\em SIAM journal on imaging sciences}~{\bf 6}(3),  1499--1530 (2013).

\bibitem{tsui2008performance}
Tsui, P.-H., Yeh, C.-K., Chang, C.-C., and Chen, W.-S., ``Performance
  evaluation of ultrasonic nakagami image in tissue characterization,'' {\em
  Ultrasonic imaging}~{\bf 30}(2),  78--94 (2008).

\bibitem{tehrani2022robust}
Tehrani, A.~K., Rosado-Mendez, I.~M., and Rivaz, H., ``Robust scatterer number
  density segmentation of ultrasound images,'' {\em IEEE Transactions on
  Ultrasonics, Ferroelectrics, and Frequency Control}~{\bf 69}(4),  1169--1180
  (2022).

\bibitem{zhang2020deep}
Zhang, L., Vishnevskiy, V., and Goksel, O., ``Deep network for scatterer
  distribution estimation for ultrasound image simulation,'' {\em IEEE
  Transactions on Ultrasonics, Ferroelectrics, and Frequency Control}  (2020).

\bibitem{Hruska2009}
Hruska, D.~P. and Oelze, M.~L., ``Improved parameter estimates based on the
  homodyned k distribution,'' {\em IEEE transactions on ultrasonics,
  ferroelectrics, and frequency control}~{\bf 56}(11),  2471--2481 (2009).

\bibitem{nam2012comparison}
Nam, K., Rosado-Mendez, I.~M., Wirtzfeld, L.~A., Ghoshal, G., Pawlicki, A.~D.,
  Madsen, E.~L., Lavarello, R.~J., Oelze, M.~L., Zagzebski, J.~A.,
  O’Brien~Jr, W.~D., et~al., ``Comparison of ultrasound attenuation and
  backscatter estimates in layered tissue-mimicking phantoms among three
  clinical scanners,'' {\em Ultrasonic imaging}~{\bf 34}(4),  209--221 (2012).

\bibitem{Rosado2016}
{Rosado-Mendez}, I.~M., {Drehfal}, L.~C., {Zagzebski}, J.~A., and {Hall},
  T.~J., ``Analysis of coherent and diffuse scattering using a reference
  phantom,'' {\em IEEE Transactions on Ultrasonics, Ferroelectrics, and
  Frequency Control}~{\bf 63}(9),  1306--1320 (2016).

\bibitem{yurtkulu2019semantic}
Yurtkulu, S.~C., {\c{S}}ahin, Y.~H., and Unal, G., ``Semantic segmentation with
  extended deeplabv3 architecture,'' in [{\em 2019 27th Signal Processing and
  Communications Applications Conference (SIU)}{\nolinebreak\hspace{0.1em}]},
  1--4, IEEE (2019).

\bibitem{liu2019variance}
Liu, L., Jiang, H., He, P., Chen, W., Liu, X., Gao, J., and Han, J., ``On the
  variance of the adaptive learning rate and beyond,'' in [{\em International
  Conference on Learning Representations}{\nolinebreak\hspace{0.1em}]},
  (2019).

\end{thebibliography}
\bibliographystyle{spiebib} 

\end{document}